 \documentclass[final,3p,times]{elsarticle}
\usepackage{amssymb}
\usepackage{sidecap}
\usepackage{caption}
\usepackage{subfigure}
\usepackage{url}
\usepackage{amstext}

\usepackage{ulem}
\usepackage{xcolor}

\renewcommand{\vec}{\mathbf}

\journal{Physica A}

\begin{document}

\begin{frontmatter}



\title{Empirical study on social groups in pedestrian evacuation dynamics}


\author[THP]{Cornelia von Kr\"uchten\corref{cor1}}
\ead{cvk@thp.uni-koeln.de}
\cortext[cor1]{Corresponding author}
\author[THP,Did]{Andreas Schadschneider}
\ead{as@thp.uni-koeln.de}

\address[THP]{Institut f\"ur Theoretische Physik, Universit\"at zu K\"oln, 
Germany}
\address[Did]{Institut f\"ur Physik und ihre Didaktik, Universit\"at zu K\"oln,
Germany}

\begin{abstract}
Pedestrian crowds often include social groups, i.e. pedestrians that walk together because of social relationships. They show characteristic configurations and influence the dynamics of the entire crowd. In order to investigate the impact of social groups on evacuations we performed an empirical study with pupils. Several evacuation runs with groups of different sizes and different interactions were performed. New group parameters are introduced which allow to describe the dynamics of the groups and the configuration of the group members quantitatively. The analysis shows a possible decrease of evacuation times for large groups due to self-ordering effects. Social groups can be approximated as ellipses that orientate along their direction of motion. Furthermore, explicitly cooperative behaviour among group members leads to a stronger aggregation of group members and an intermittent way of evacuation.
\end{abstract}


\end{frontmatter}


%\textbf{Highlights}
%\begin{itemize}
%\item Quantitative analysis of an empirical study on the influence of social groups in evacuation scenarios
%\item Increasing the social group size can have an positive impact on evacuation times due to self-ordering phenomena
%\item Introduction of new quantitative group parameters considering the groups' centre of mass and its dynamics and the shape of the groups
%\item Social groups order along their direction of motion
%\end{itemize}

\section{Introduction}
\label{S1:Introduction}

The significance of social groups in pedestrian crowds is known for
almost forty years. In 1977, Aveni \cite{Aveni1977} found that most
pedestrians do not walk alone, but in pairs or groups. Likewise, more
recent studies observed the dominance of groups in crowds of
pedestrians \cite{Reuter2014,Schultz2014,Xu2010}. Moussaid et al. \cite{Moussaid2010} reported that up to 70\% of urban pedestrian
traffic happens in social groups, whereas 95\% of pedestrians
were found to walk in groups at major events \cite{Oberhagemann2014}.

In this contribution, the term ``crowd" describes the entire group of walking or evacuating people. Groups of pedestrians that walk or stand together because of social
relationships and interactions are referred to as ``social groups"
(see \cite{Moussaid2010} and references therein). Using these terms, we do not want to imply collective behaviour as it is mentioned in \cite{Templeton2015}. The presence of social groups can
influence the dynamics of pedestrian crowds. Social groups often move
slower and reduce the walking speed of the pedestrians
\cite{Schultz2014,Gorrini2014,Gorrini2016,Wei2014}. However, Manenti et al.
\cite{Manenti2010} observed that group members can walk faster than individuals in high density scenarios. The average walking speed decreases with increasing group size \cite{Schultz2014,Moussaid2010,Oberhagemann2014,Koester2011}.

Pedestrians of social groups order in certain configurations depending
on the surrounding density and the group size. Pairs and larger groups
at lower density walk abreast. At higher densities, groups of three
pedestrians often walk in a ``V"-like shape. This configuration
becomes ``U"-like for groups with four pedestrians. Larger groups tend
to split up to smaller groups of two or three members
\cite{Schultz2014,Moussaid2010,Costa2010,Xi2014,Zanlungo2013}. In
pedestrian crowds the averaged minimum distance headway within a
social group increases with increasing group size. This leads to a
larger space requirement for larger groups \cite{Duives2014}.

Based on simulation results Reuter et al. \cite{Reuter2014} identify
large social groups as ``moving obstacles" amongst pedestrians. As a
consequence, the authors assume that large cohesive groups inhibit
fast evacuations. K\"oster et al. \cite{Koester2011} performed an
experimental study with students. They had to egress from their
classroom starting at their desks and enter the room again afterwards.
The authors of the study observed a negative impact of groups on
egress times, but a positive impact on ingress times. The students
entered the room starting in the same configuration of the crowd as it
results from the egress process. Thereby, the separation of the participants
into social groups caused the ordering of students in accordance with their position in the classroom before the ingress process started.
The authors assume that this kind of ordering effect was responsible
for the faster ingress with groups. A more recent study
\cite{Guo2015} also found that pairs can evacuate faster than
individuals. Laboratory experiments performed by Bode et. al \cite{Bode2015} revealed that social groups of three evacuated slower compared to individuals due to a larger pre-movement time and larger time to reach the vicinity of the target.

In order to investigate social groups in pedestrian crowds further,
this contribution presents the first results of laboratory experiments
on social groups in evacuation scenarios. These experiments were part
of a joint study by the Universities of Cologne and Wuppertal and the
Forschungszentrum J\"ulich which aimed at the investigation of the
influence of inhomogeneities on the fundamental diagram
\cite{Wtalinprep} and evacuation times. Here we report preliminary results
from experiments on evacuation scenarios with pairs and larger social
groups.

%%%%%%%%%%%%%%%%%%%%%%%%%%%%%%%%%%%%%%%%%%%%%%%%%%%%%%%%%%%%%%%%%%%%%%%%

\section{Experimental study}
\label{S2:ExpStudy}

The experiments were performed in November 2015 and April 2015 in two
schools in Wuppertal, Germany (Gymnasium Bayreuther Stra{\ss}e,
``GymBay", and Wilhelm-D\"orpfeld-Gymnasium, ``WDG"). The pupils of
different classes participated as part of project work (see also
\cite{TGF15, PED16}).

\subsection{Experimental concept}
\label{SS2.1:ExpConcept}

The experiments consisted of several evacuation runs. The focus was on
varying the composition of the crowd of pupils. Different
configurations helped to investigate the influence of different
parameters. Overall, there were three quantities that specified the
composition of the crowd:
\begin{itemize}
\item \textit{Social group composition}: There were two classes of age: children (aged around 11 years) and young adults (aged around 16 years).
The groups could therefore consist only of children, only of youths and of mixtures of both. A distribution of body height of the runs that are used in the analysis can be found in \ref{ASS1.2:BodyHeight}.
\item \textit{Social group interaction}: The interaction between group
  members was specified in two ways:
        \begin{itemize}
        \item \textit{Bond between group members}: Members of the same
          social group could be connected either loosely or fixed.
          Loosely bonded groups had to try to stay together through
          eye contact, fixed bonded group members had to hold physical
          contact, e.g. hold each others' hands.
        \item \textit{Hierarchy of group members}: The hierarchy
          between group members could be flat, with each pupil treated
          equally. They had to leave the room as well as to stay
          together in their social group. Otherwise, one group member
          was declared as `leader', the other ones as `followers'.
          Leaders had to leave the room without regarding other
          members of the groups. Followers had to stay as close as possible to their leader.
        \end{itemize}
      \item \textit{Social group size}: The pupils evacuated individually, in
        pairs or in larger groups with four, six or eight participants
        per group.
\end{itemize}
Large groups were built successively. Two pupils were teamed up to
pairs. Groups of four consisted of two pairs, groups with six
participants of three pairs and groups of eight were built by four
pairs (two groups of four). Members of larger groups were bound only
loosely. If required, the leader was chosen randomly in age-matched
pairs. In pairs or groups of mixed ages the older pupil or one of the
older pupils were assigned as leader. For age-matched large groups
there was no leader at all.

%%%%%%%%%%%%%%%%%%%%%%%%%%%%%%%%%%%%%%%%%%%%%%%%%%%%%%%
\subsection{Experimental set-up}
\label{SS2.2:ExpSetup}

\begin{SCfigure}[0.7][ht]
\centering
\includegraphics[scale=0.18]{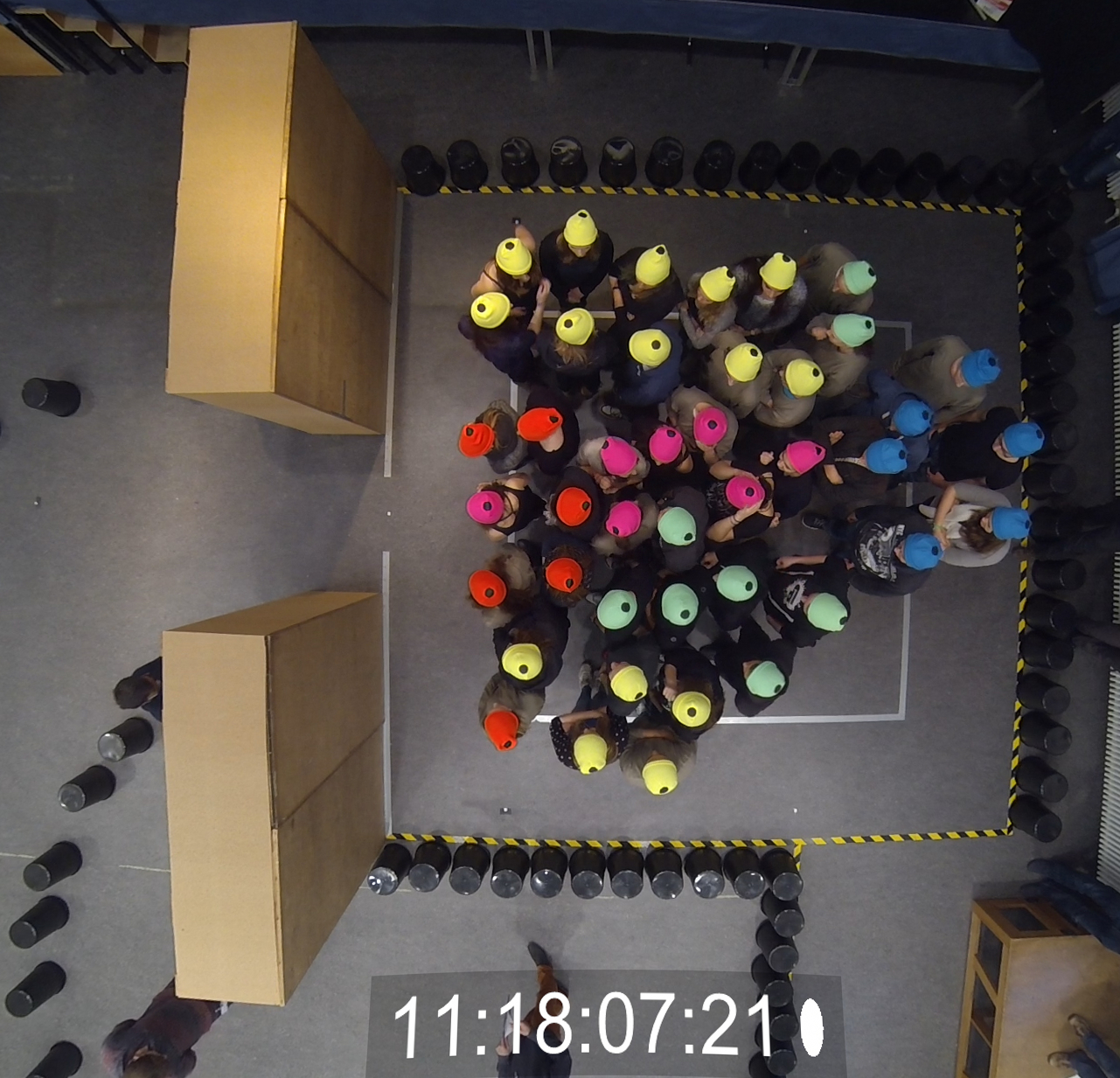}
\caption{This screenshot from video recordings shows the 
  experimental set-up which consisted of a square area
  (right) with two artificial walls (left) that formed the exit
  door. The students assembled in the square starting area in the
  centre and left the room using the door. They walked down the aisle
  and waited in front of the entrance at the bottom of the picture.
  The colour of the caps does not represent the different social
  groups, but the body height of the pedestrian.}
\label{Fig1:Raum}
\end{SCfigure}

The experimental set-up consisted of a rudimentary room built in the
schools' assembly hall. This room was a square area of 5$\times$5
m$^2$ bounded by small buckets on three sides (see
Fig.~\ref{Fig1:Raum}). There was a gap on one side that was used as an
entrance to the experimental area. The fourth side was an artificial wall with an exit door of variable width. Door widths of 0.8~m and 1.2~m were used. Behind the exit door more buckets formed an aisle
that led to a waiting area in front of the entrance. In the centre of
the experimental room a square starting area of 3$\times$3~m$^2$ was
marked. To record the experiments a video camera system was mounted on
the ceiling of the hall.  All pupils wore caps with different colours.
Each colour represented a certain interval of body heights which was
measured for each pupil before the experiments. They were used for an
exact determination of the pupils' position by reducing perspective errors \cite{Boltes2010}. In addition, the caps came
with black points in the middle of the head. This point was used for
the recognition and tracking of the pupils in the videos
\cite{Boltes2010}. The video recordings were processed using the
\textsc{PeTrack} software \cite{Boltes2010,BoltesDiss} so that
trajectories for each pupil in each run were available for our
analysis. All videos and trajectories are available under \url{http://ped.fz-juelich.de/database}.

%%%%%%%%%%%%%%%%%%%%%%%%%%%%%%%%
% Suggested size: Single
%%%%%%%%%%%%%%%%%%%%%%%%%%%%%%%%

%%%%%%%%%%%%%%%%%%%%%%%%%%%%%%%%%%%%%%%%%%%%%%%%%%%%%%%
\subsection{Experimental procedure}
\label{SS2.3:ExpProcedure}

In each experimental run, there were 32 - 46 participants. In the waiting area
in front of the entrance the pupils were told in which configuration
they had to evacuate and were separated into pairs or groups, if
required.  Afterwards, they entered the room and assembled in the
starting area, uniformly distributed. Groups or pairs should stand
together before the start of each experimental run. The students were told to raise their hands in this groups so that single social groups could be identified in the video recordings later. After an acoustic starting signal the pupils left the room.
They were told to go briskly but not to run or to scramble. The
current group configuration had to maintain during the whole egress.
After passing the exit door the pupils had to walk down the aisle and
to assemble again in the waiting area.

%%%%%%%%%%%%%%%%%%%%%%%%%%%%%%%%%%%%%%%%%%%%%%%%%%%%%%%%%%%%%%%%%%%%%%%%

\section{Analysis of experimental data}
\label{S3:Analysis}

In order to investigate the influence of groups the data of the youths
of both schools were used. We compare runs with large groups (four,
six and eight participants), loosely bonded pairs without
leader-follower relationship and, if available, individual runs. In
addition, the youths of ``GymBay" performed a run with groups of six
and explicit cooperative behaviour. They were told to particularly
concentrate on their group members. A detailed list of the used runs can be found in the appendix.

%%%%%%%%%%%%%%%%%%%%%%%%%%%%%%%%%%%%%%%%%%%%%%%%%%%%%%%

\subsection{Macroscopic analysis}
\label{SS3.1:MacroAnalysis}

A first approach to consider social groups in pedestrian evacuation
dynamics is to investigate their influence on the entire crowd.
Therefore, the macroscopic quantities of evacuation times, density
distributions and information from spatio-temporal diagrams can be
regarded. They may give some indication of changes of the crowd's dynamic due to the presence of social groups.

\subsubsection{Evacuation times}
\label{SSS3.1.1:EvacTimes}

\begin{figure}[h]
\centering
\includegraphics[scale=0.37]{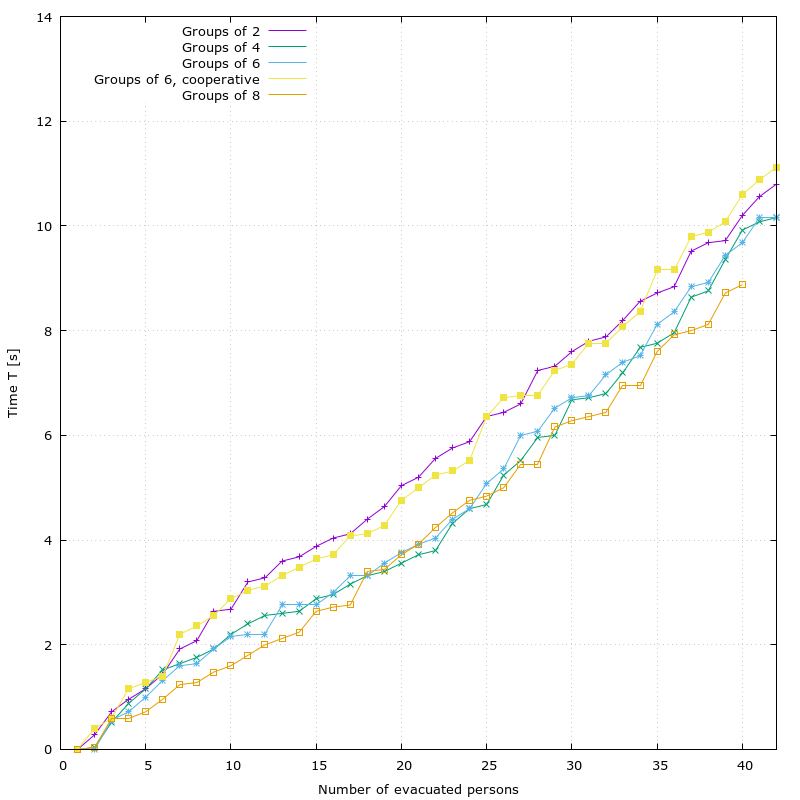}
\caption{The evacuation time plot for ``GymBay" shows an early 
  separation of the curves. The runs with groups of four, six and
  eight are significantly faster than the pairwise run and the run
  with cooperative behaviour.}
\label{Fig2:EZGymBay}
\end{figure}

\begin{figure}[h]
\centering
\includegraphics[scale=0.37]{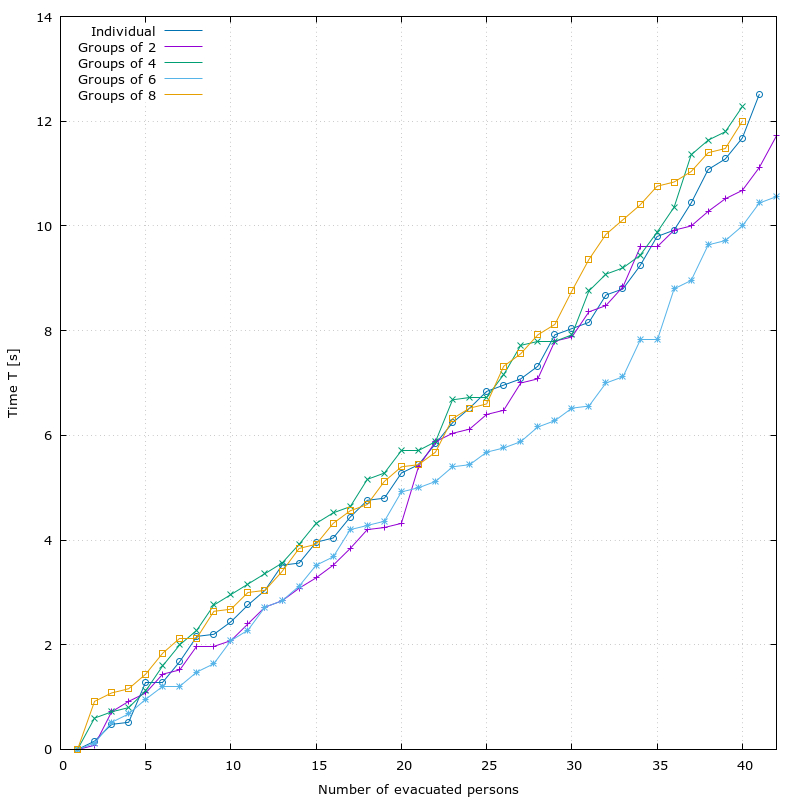}
\caption{Evacuation time plot for ``WDG": The evacuation time curves 
  do not separate, only the run with groups of six becomes faster
  after approx.\ 20 evacuated persons.}
\label{Fig3:EZWDG}
\end{figure}

Figures~\ref{Fig2:EZGymBay} and~\ref{Fig3:EZWDG} show the number of
evacuated persons plotted against the respective evacuation times for
each school, respectively. The evacuation time of each person can be
calculated as the time difference between the beginning of the
evacuation and the moment when the participant passes the door. Since
the acoustic starting signal is not audible on the video recordings,
the beginning of the evacuation can be determined only inaccurately.
To compensate this and other delays (e.g. pre-movement time, delays
due to a lack of concentration) the evacuation time of the first
evacuated person is subtracted as an off-set from all the other times.
Additionally, the adjusted times allows for a better comparison
between different runs.

%%%%%%%%%%%%%%%%%%%%%%%%%%%%%%%%
% Suggested size: 1.5
%%%%%%%%%%%%%%%%%%%%%%%%%%%%%%%%
%\begin{SCfigure}[0.3][ht]
%\centering
%\includegraphics[scale=0.4]{Figures/Fig2-EZGymBay.png}
%\caption{The evacuation time plot for ``GymBay" shows an early splitting of the curve. The runs with groups of four, six and eight are significantly faster than the pairwise run and the run with cooperative behaviour.}
%\label{Fig2:EZGymBay}
%\end{SCfigure}
%%%%%%%%%%%%%%%%%%%%%%%%%%%%%%%%

Both plots show approximately linear behaviour of the curves. The
evacuation times of ``GymBay" separate into two branches after about
three to six evacuated persons. The runs with groups of four, six and
eight persons are significantly faster than the runs with pairs and
with groups of six and cooperative behaviour. One should notice that
the run with cooperative behaviour is considerably slower than the
evacuation with the same group size but normal behaviour. There
is no such separation in the plot of ``WDG", however, the run with
groups of six becomes faster than the others in the second half of the
evacuation.  On this basis, there are two conjectures that can be
made:
\begin{enumerate}
\item   The presence of groups is not necessarily disadvantageous, in some
                cases the evacuation can be even faster with social groups.
\item   Explicitly cooperative behaviour can inhibit an advantageous
                impact of groups on the evacuation process.
\end{enumerate}

%%%%%%%%%%%%%%%%%%%%%%%%%%%%%%%%
% Suggested size: 1.5
%%%%%%%%%%%%%%%%%%%%%%%%%%%%%%%%

%\begin{SCfigure}[0.3][ht]
%\centering
%\includegraphics[scale=0.4]{Figures/Fig3-EZWDG.png}
%\caption{The evacuation times of ``WDG" do not split. Only the run with groups of six becomes faster after $\sim$ 20 evacuated persons.}
%\label{Fig3:EZWDG}
%\end{SCfigure}
%%%%%%%%%%%%%%%%%%%%%%%%%%%%%%%%

Further investigation of the pedestrians' behaviour during the evacuation
may give a hint of possible reasons for the differences in evacuation
times.

\subsubsection{Density distributions}
\label{SSS3.1.2:DensDistribution}

\begin{figure}[h]
\centering
\subfigure[Pairs]{\includegraphics[scale=0.35]{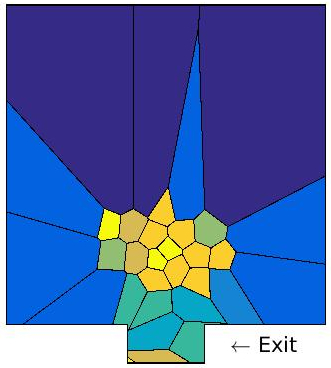}}
\subfigure[Groups of four]{\includegraphics[scale=0.35]{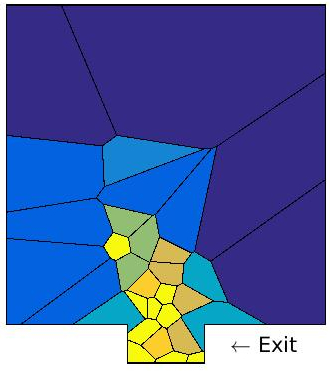}}
\subfigure[Groups of eight]{\includegraphics[scale=0.35]{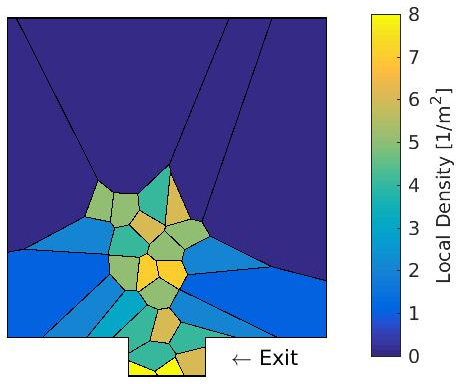}}
\\
\subfigure[Pairs]{\includegraphics[scale=0.35]{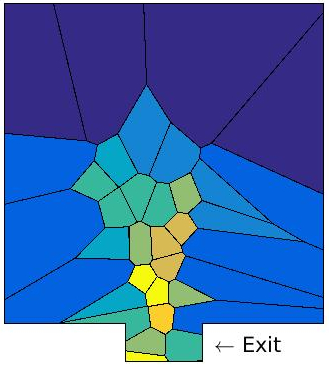}}
\subfigure[Groups of six]{\includegraphics[scale=0.35]{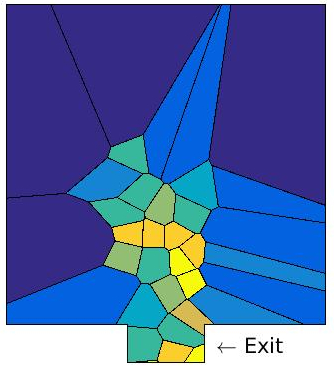}}
\subfigure[Groups of eight]{\includegraphics[scale=0.35]{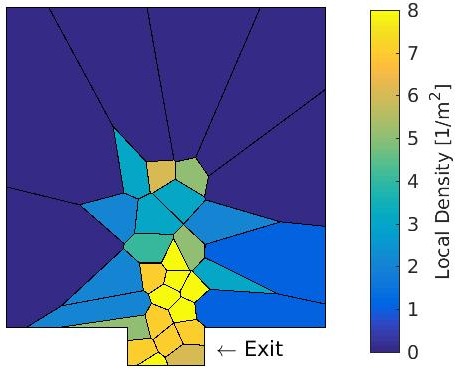}}
\caption{Exemplary density distributions of ``GymBay" ((a)-(c)) 
  and ``WDG" ((d)-(f)) at $T=4$ s. For ``GymBay", there is a rotund
  shape of the whole group in the run with pairs and an elongated
  configuration for larger social groups. All runs show queue-like
  configurations in front of the door for ``WDG".}
\label{Fig4:DD}
\end{figure}

By means of density distributions the configuration of the pedestrians
during the evacuation can be depicted. A density profile can be
determined using Voronoi diagrams \cite{Voronoi1907,Steffen2010}.
These diagrams consist of Voronoi cells for each pedestrian. Each cell
represents the ``personal space" \cite{Liddle2011} of the respective person. The size of the Voronoi cells
can be used as a measure of density: the (local) density encountered
by a person is inversely proportional to the area of the corresponding
Voronoi cell \cite{Steffen2010}. In Fig.~\ref{Fig4:DD} the cells are
coloured corresponding to their size: large cells (low density) are
coloured in shades of blue, smaller cells (high density) in shades of
red. In doing so, dense configurations of pedestrians can be
recognized easily.

Density distributions of different runs for ``GymBay" and ``WDG" are
exemplarily shown in Fig.~\ref{Fig4:DD}. The participants congest in front
of the exit. The pupils in the pairwise run for ``GymBay'' order in a
broad shape symmetrically around the door. This configuration is
elongated for the runs with larger social group sizes. The pupils order
rather behind one another than next to each other when evacuating in larger
groups. This queue-like formation of the crowd is observed in all runs
of ``WDG" independently of the group size.
%%%%%%%%%%%%%%%%%%%%%%%%%%%%%%%%
% Suggested size: 1.5
%%%%%%%%%%%%%%%%%%%%%%%%%%%%%%%%

Nevertheless, the differences in the positioning of the pedestrians of
the ``GymBay" experiments may explain the differences in evacuation
times. The run with pairs was significantly slower than the runs with
groups of larger size and normal behaviour. A kind of self-ordering
phenomenon within the groups may lead to an ordering effect
of the entire crowd. In front of the exit, the
pedestrians compete for space in the door. The observed ordering
effect could reduce the number of these conflicts. As a consequence,
the evacuation processes smoother and faster. There
are no distinct differences in evacuation times for ``WDG", except for
the run with groups of six at the end, but there are also no
differences in the pupils' configuration. Also the density
distribution does not supply an explanation for the slower evacuation
with cooperative behaviour. Therefore, one
has to chose other approaches.

\subsubsection{Spatio-temporal diagrams}
\label{SSS3.1.3:SpatTempDiagram}

\begin{figure}[h]
\centering
\subfigure[Groups of six]{\includegraphics[scale=0.263]{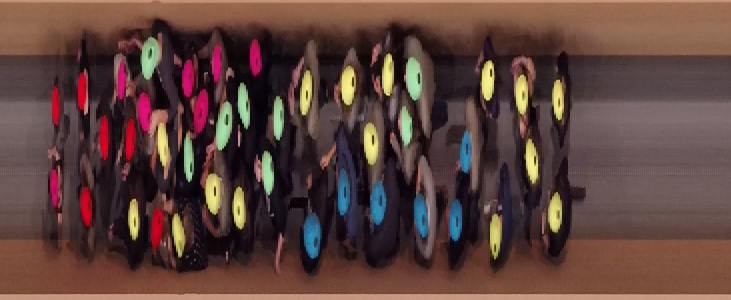}}
\subfigure[Groups of six, cooperative]{\includegraphics[scale=0.263]{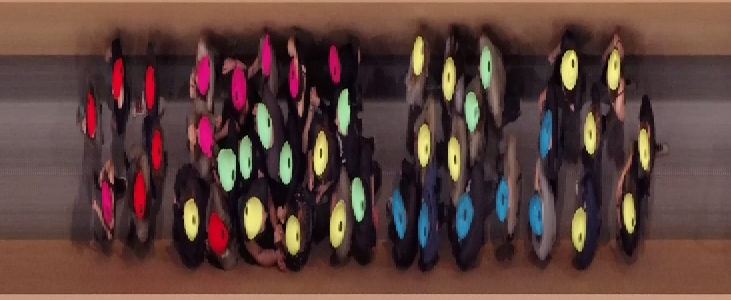}}
\subfigure[Pairs]{\includegraphics[scale=0.263]{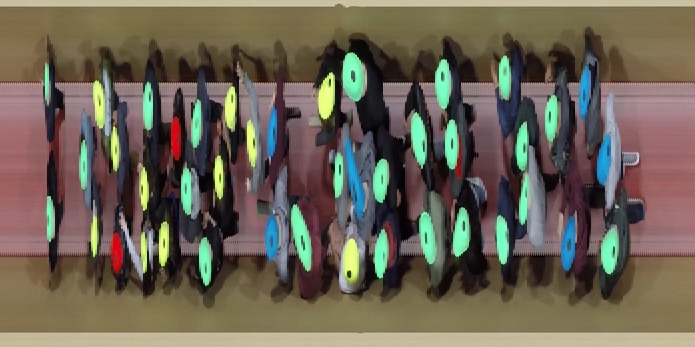}}
\subfigure[Groups of six]{\includegraphics[scale=0.263]{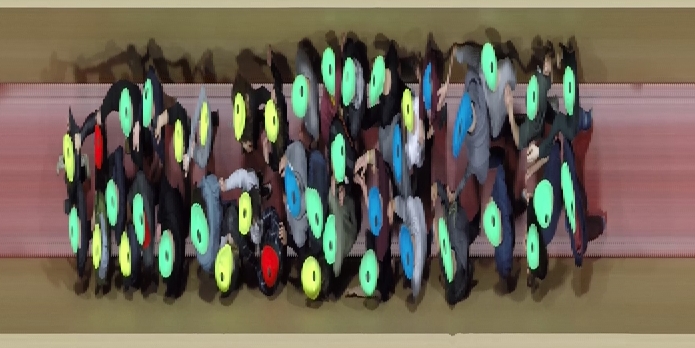}}
\caption{Spatio-temporal diagrams for ``GymBay" (above) and ``WDG" 
  (below). The pupils of ``GymBay" move in a compact way in groups
  with normal behaviour, but in an intermittent way with cooperative
  behaviour. For ``WDG", the run with groups of six shows a more
  compact crowd than the other runs. The colour of the caps represents
  the body height, not the social group.}
\label{Fig5:STD}
\end{figure}

The compactness of the pedestrians during the evacuation can be visualized
using spatio-temporal diagrams \cite{Zuriguel2014,Boltes2014}. They
represent the temporal development of a certain spatial region. In the
case of the evacuation experiments the line of pixels corresponding to
the end of the exit door built by the platforms is plotted for every
frame (Fig.~\ref{Fig5:STD}).
%%%%%%%%%%%%%%%%%%%%%%%%%%%%%%%%
% Suggested size: 1.5
%%%%%%%%%%%%%%%%%%%%%%%%%%%%%%%%

In all runs with normal behaviour for ``GymBay" as well as for ``WDG"
the pupils move in a compact way. There are no distinct gaps between
single pedestrians over the course of time. In contrast, the (slower)
run with groups of six and cooperative behaviour shows an intermittent
behaviour, see Fig.~\ref{Fig5:STD}~(b). The students are separated into
four bursts. Each burst mostly consists of closed social groups.
Cooperative behaviour seems to lead to a stronger aggregation to the
own group members and a neglect of other pupils at the same time. This
loss of cooperation with the rest of the students
may slow down the evacuation process.

The pupils of ``WDG" evacuate in a continuous way for all group sizes.
They are distributed uniformly, but loosely, except for the run with
groups of six. In this case, the pedestrian flow is higher. This is
represented by a higher density in the spatio-temporal diagram
Fig.~\ref{Fig5:STD}~(d). It is conceivable that this very compact
configuration accelerates the evacuation process. The runs with normal
behaviour of ``GymBay" also show temporal periods of higher flow. In
case of larger groups, this period occurs in the first third of the
evacuation. For the pairwise run a period of higher flow can be
observed in the middle of the evacuation. It might be advantageous if
there is a high pedestrian flow at an early state of the evacuation.

%%%%%%%%%%%%%%%%%%%%%%%%%%%%%%%%%%%%%%%%%%%%%%%%%%%%%%%
\subsection{Microscopic analysis}
\label{SS3.2:MicroAnalysis}

Besides the influence of the social groups on the entire crowd the
behaviour and dynamics of the groups themselves are interesting. In
the following sections several ``group parameters" are introduced. The
velocity of the centre of mass helps to describe the dynamics of a
group. Furthermore, the shape and
elongation of a group provides information about the positioning of
the group members. The orientation of the group is also regarded on
different ways. For a more detailed description, we refer to \cite{vonKruechten2016}.

\subsubsection{Centre of mass of social groups}
\label{SSS3.2.1:CoM}

The dynamics of a social group as whole is described by the behaviour
of its centre of mass. This approach is adopted from the investigation
of starling flocks \cite{Ballerini2008}. The position of the centre of
mass is calculated as the averaged position of all group members:

\begin{equation}
\vec{R}_\text{CoM}(t)=\frac{1}{N_m}\sum_{i=1}^{N_m}\vec{r}_i(t)
\end{equation}

In order to describe the dynamics we consider the velocity of the centre of mass
\begin{equation}
\vec{v}_\text{CoM}(t)=\frac{\vec{R}_\text{CoM}(t)-\vec{R}_\text{CoM}(t-\Delta t)}{\Delta t}
\end{equation}
that describes the change of position of the centre of mass with time. In this case, $\Delta t$ corresponds to the frame rate of the video recordings, $\Delta t = \frac{1}{25}$~s.

\begin{figure}
\centering
\includegraphics[width=0.7\textwidth]{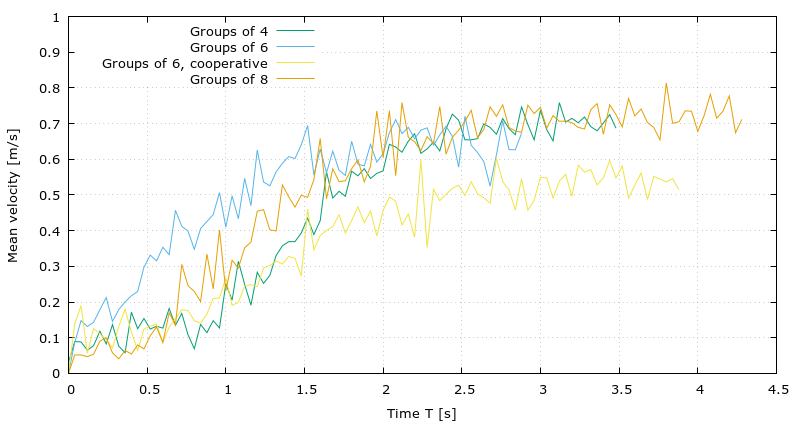}
\includegraphics[width=0.7\textwidth]{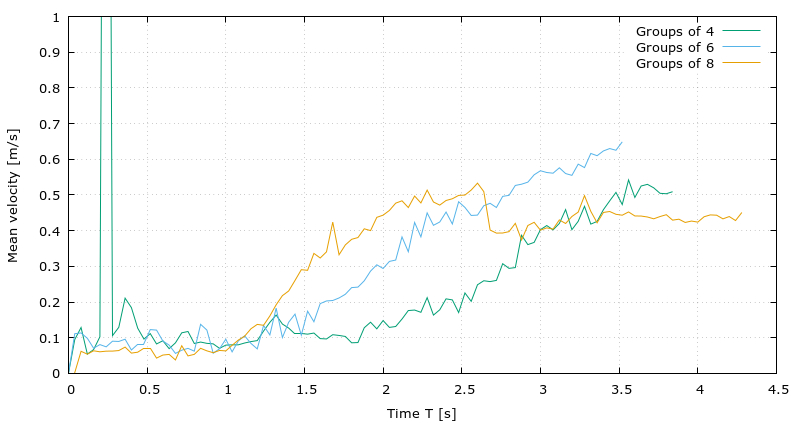}
%\subfigure[``GymBay"]{\includegraphics[width=0.49\textwidth]{Figures/Fig6-MeanVel-GymBay-oT.png}}
%\subfigure[``WDG"]{\includegraphics[width=0.49\textwidth]{Figures/Fig7-MeanVel-WDG-oT.png}}
\caption{Mean velocity of the centre of mass. For ``GymBay" (top) the 
  velocity increases and then reaches a constant level. This level is
  smaller for cooperative behaviour than for normal behaviour. For
  ``WDG" (bottom) the single velocity curves are more diverse.}
\label{Fig6:MeanVel}
\end{figure}

Fig.~\ref{Fig6:MeanVel} shows velocity of the centre of mass at time $t$ averaged over the number of
social groups for ``GymBay" and ``WDG". The calculation of the mean
velocity is cancelled as soon as one social group has left the region
of detection.\\
%%%%%%%%%%%%%%%%%%%%%%%%%%%%%%%%
% Suggested size: 1.5 or 2 column
%%%%%%%%%%%%%%%%%%%%%%%%%%%%%%%%
All velocity curves for ``GymBay" increase almost linearly at the
beginning of the evacuation and flatten after approximately 1.5~s. The
groups with normal behaviour reach a nearly similar constant velocity
around 0.7~ms$^{-1}$. In contrast, the run with cooperative behaviour
is significant slower. This curves bottoms out at approximately
0.5~ms$^{-1}$. This difference in velocity may be an additional
explanation for the longer evacuation time for the run with
cooperative groups. Overall, in all runs the groups accelerate and
reach a nearly constant velocity in the first few seconds of
evacuation.

For the runs of ``WDG", the velocity profile is different. The groups
do not accelerate in the first second. The increase is almost linear
only for the run with groups of six. This curve also does not flatten
during the measurement period. The runs with groups of four and eight
participants show different increases at the beginning. The velocity
of the groups of eight reaches a constant level at above 0.4~m/s, the
curve of the run with groups of four flattens at a slightly higher
velocity. These various progresses of the mean velocity are not
explainable at first sight. For whatever reasons the dynamics of the
groups at ``WDG" was different from that of ``GymBay" and more
inconsistent.

\subsubsection{Shape of social groups}
\label{SSS3.2.2:Shape}

\begin{figure}[h]
\centering
\subfigure[Example ellipse]{\includegraphics[scale=0.075]{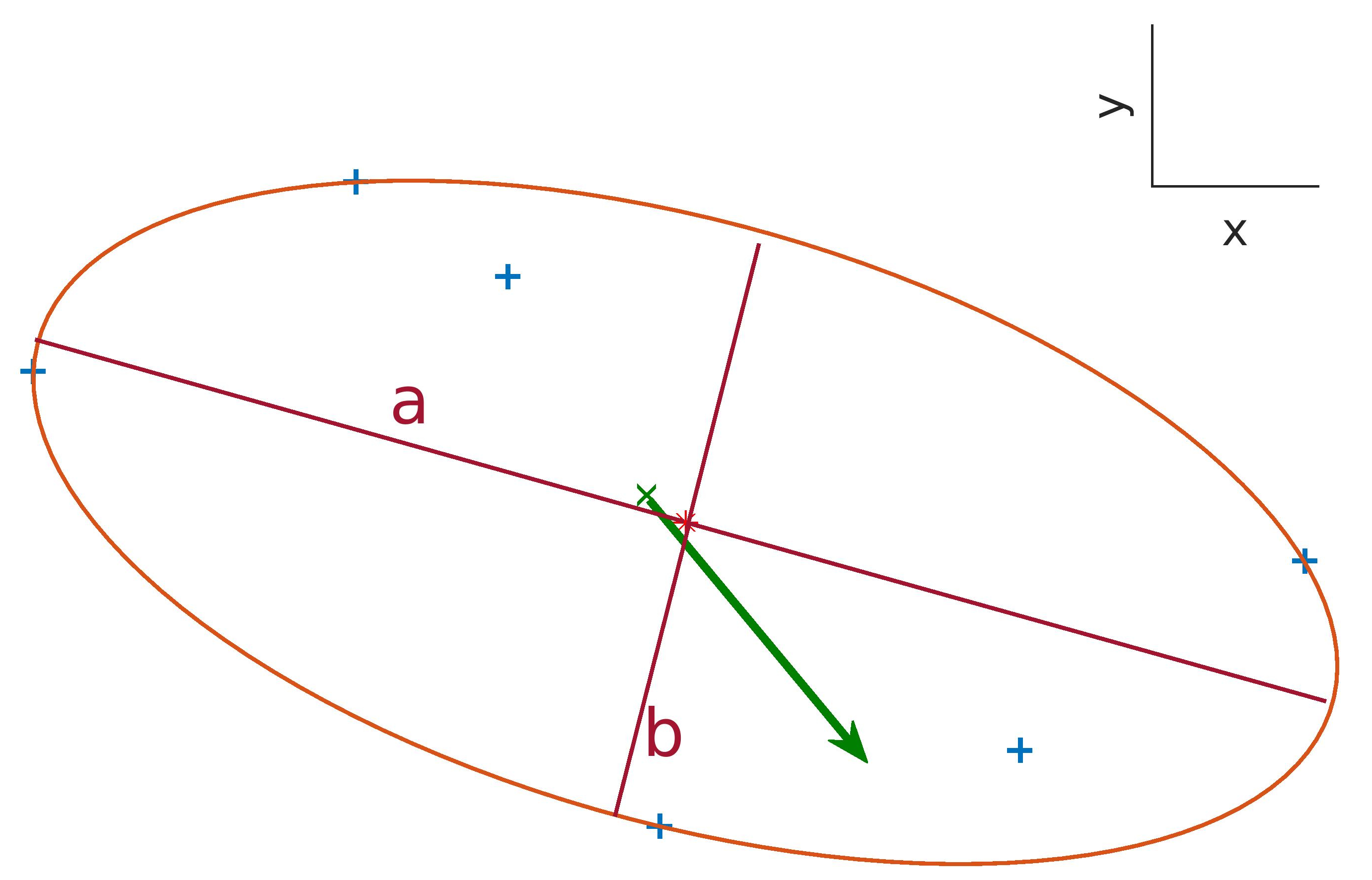}}
\hspace{1cm}
\subfigure[Definition of the CoM orientation]{\includegraphics[scale=0.3]{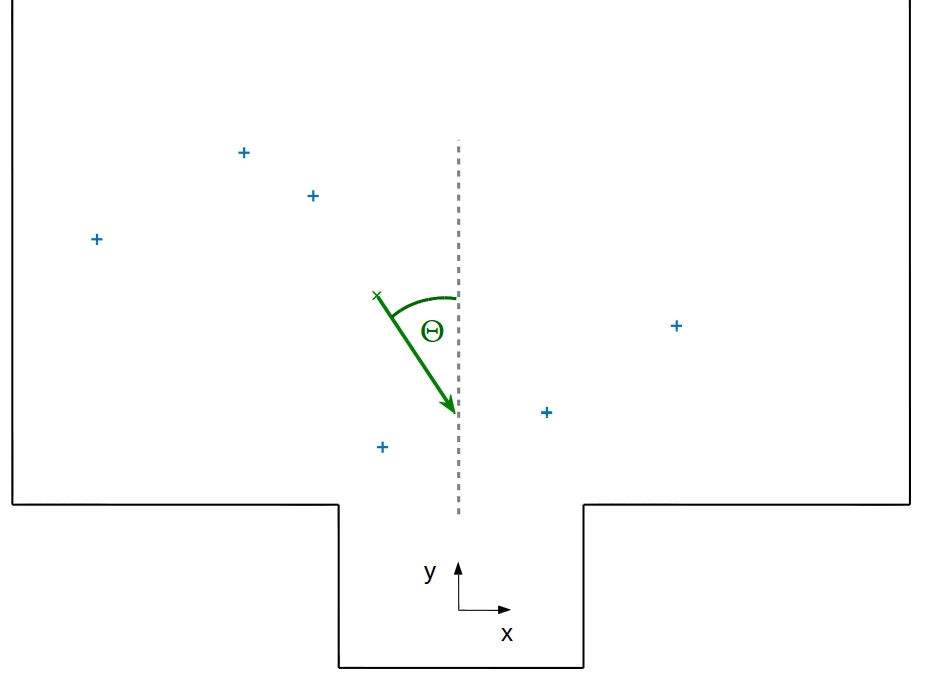}}
\caption{(a) Example ellipse of a social group with major and minor axis $a$ and $b$, respectively. The blue symbols (+) indicate the position of the members of the social groups. The green cross (x) denotes the position of the centre of mass. The green arrow depicts an exemplary velocity vector of the centre of mass. (b) $\Theta$ is the angle characterising the orientation of the centre of mass as defined in (3).}
\label{Fig6.5:EllipseBeispiel}
\end{figure}

Besides the dynamics of the group as a
  whole, the behaviour and configuration of the pupils within the
social group is also of interest. The shape of the group can be
approximated as an ellipse. This ellipse should incluce all group members and have the minimal possible area at the same time. Therefore, it is directly determined by the position of the pedestrians within the social group. Even its aspect ratio, i.e. the ratio of
its major and minor axis, and its area is determined by the
positioning of the group members \cite{Moshtag2006,Moshtag2007}. An exemplary ellipse is shown in Fig. \ref{Fig6.5:EllipseBeispiel} (a).

%%%%%%%%%%%%%%%%%%%%%%%%%%%%%%%%
% Suggested size: Single
%\begin{SCfigure}[0.7][ht]
%\centering
%\includegraphics[scale=0.075]{Figures/E1-Ellipse-GymBay6erGr5-11-3.jpg}
%\caption{\add{Exemplary ellipse for a social groups of six participants in the experimental room. The group is approximated as an ellipse that contains all group members (in red) and that has the minimal area. The elongation, area and elongation is directly determined by the positioning of the pedestrians (in blue) within the social groups. Therefore, these quanitites can be used to investigate the behaviour of the single social group during its movement.}}
%\label{Fig6.5:EllipseBeispiel}
%\end{SCfigure}
%%%%%%%%%%%%%%%%%%%%%%%%%%%%%%%%

%%%%%%%%%%%%%%%%%%%%%%%%%%%%%%%%

% Suggested size: 1.5 or 2 column
\begin{figure}[h]
\centering
\includegraphics[width=0.6\textwidth]{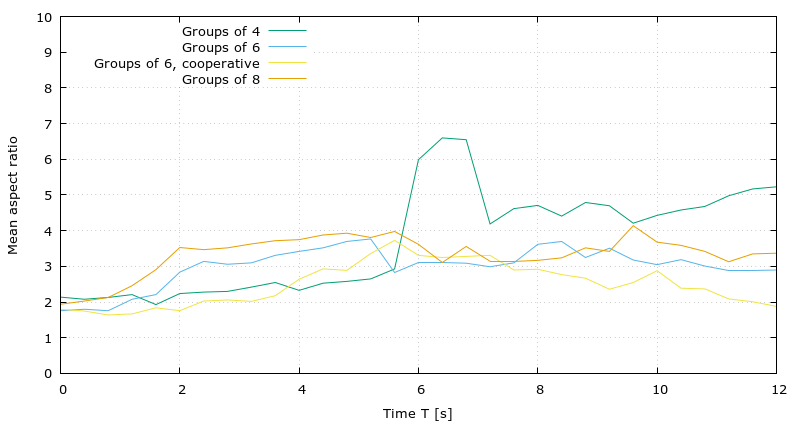}
\includegraphics[width=0.6\textwidth]{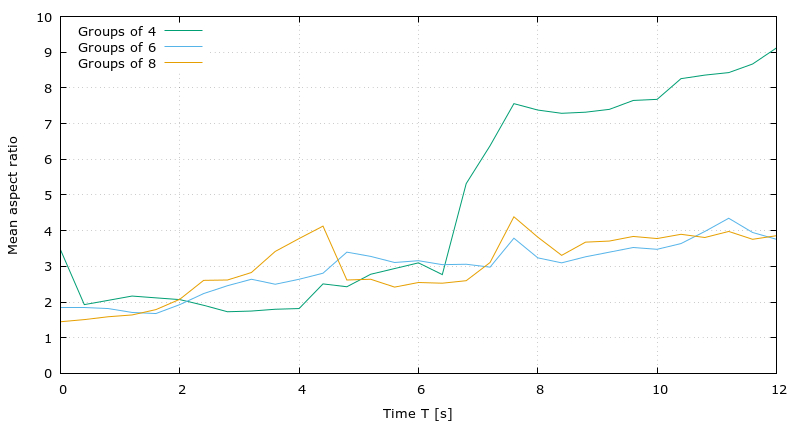}
%\subfigure[``GymBay"]{\includegraphics[width=0.49\textwidth]{Figures/Fig8-MeanARGymBay-oT.png}}
%\subfigure[``WDG"]{\includegraphics[width=0.49\textwidth]{Figures/Fig9-MeanARWDG-oT.png}}
\caption{Mean aspect ratio of the ellipse for ``GymBay" (top) and 
  ``WDG" (bottom). The groups form prolately elongated ellipses into
  the direction of the door.}
\label{Fig7:MeanAspRat}
\end{figure}
%%%%%%%%%%%%%%%%%%%%%%%%%%%%%%%%

The aspect ratio averaged over the number of groups is plotted for
both schools in Fig.~\ref{Fig7:MeanAspRat}. For all runs the aspect
ratio is distinctly greater than one. It indicates that the groups
form a prolate ellipse. In this case the pedestrians have a elongated
configuration in the direction of the exit door. The aspect ratio
increases with time for all runs, only for ``GymBay" it decreases
again later for most of the runs. It is remarkable that the aspect
ratio of the runs with groups of six and eight develop very similarly
with time in both schools. The aspect ratio of the runs with groups of
four increases sharply in the middle of the evacuation in both cases.
This is caused by single social groups whose members are separated
during the evacuation process. The run with cooperative behaviour
shows a similar process of the aspect ratio than the run with groups
of six and eight. However, at most times the absolute value of the
aspect ratio is smaller than in the runs with normal behaviour.
Cooperative behaviour seems to lead to a more rotund configuration of
the group members than evacuations without this instruction.

\begin{figure}[h]
\centering
\includegraphics[width=0.6\textwidth]{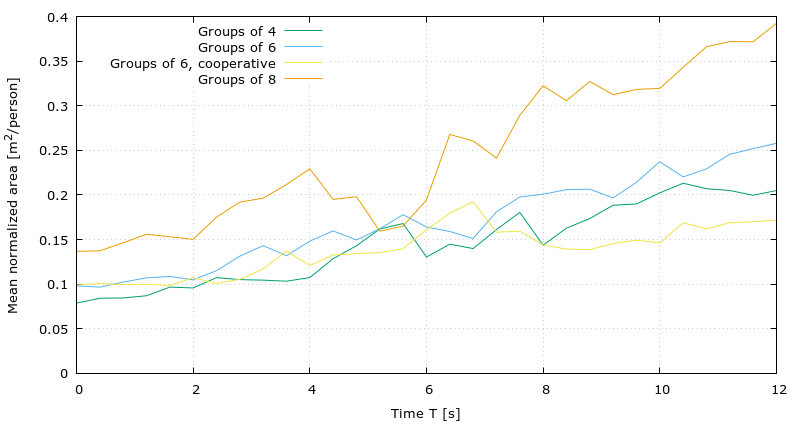}
\includegraphics[width=0.6\textwidth]{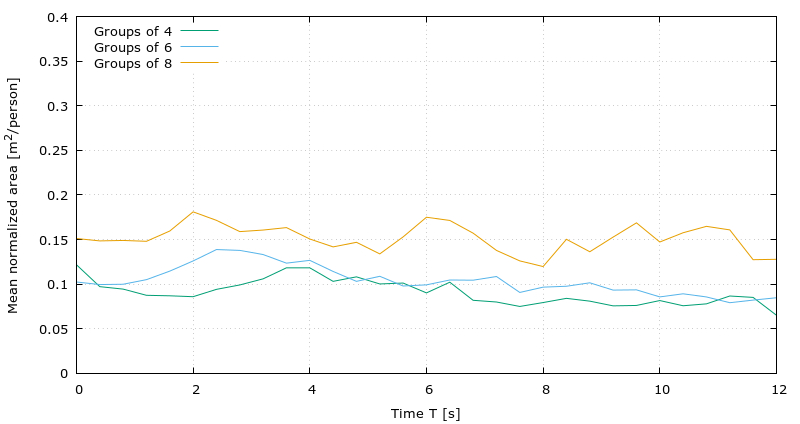}
%\subfigure[``GymBay"]{\includegraphics[width=0.49\textwidth]{Figures/Fig10-MeanNAGymBay-oT.png}}
%\subfigure[``WDG"]{\includegraphics[width=0.49\textwidth]{Figures/Fig11-MeanNAWDG-oT.png}}
\caption{Mean normalized area of the ellipse for ``GymBay" (top) 
  and ``WDG" (bottom). The space requirement per person increases with
  group size.}
\label{Fig8:MeanNormArea}
\end{figure}

The area of the ellipse can also be considered. It is normalized with
respect to the number of pedestrians of the group in order to
allow for a quantitative comparison. In doing so, the space requirement
per person within the group is determined. This approach is similar to a method that is used to probe the dispersion of social groups, e.g. in \cite{Bandini2011, Vizzari2013}. In our case the ellipse is used instead of a polygon to determine the space that is occupied by the group members. The averaged normalized
area of the ellipses for both data sets is shown in
Fig.~\ref{Fig8:MeanNormArea}. The general behaviour of the curves is
different in ``GymBay" and ``WDG". For the runs of ``GymBay" the mean
normalized area increases during the evacuation for all runs. The
group members fan out over a larger area with time. In contrast, the
normalized area decreases during the evacuation process for all group
sizes in ``WDG". In these cases the pedestrians come closer within the
groups. However, both types of behaviour are consistent within one
data set.

%%%%%%%%%%%%%%%%%%%%%%%%%%%%%%%%
% Suggested size: 1.5 or 2 column
%%%%%%%%%%%%%%%%%%%%%%%%%%%%%%%%

For both schools the normalized area is slightly larger or even
similar for groups of six compared to groups of four. In contrast, the
values for the groups of eight are significantly higher, expect for a
small outlier for ``GymBay". The run with groups of six and
cooperative behaviour shows the same normalized area than the group
with same group size but normal behaviour in the beginning of the
evacuation. The process of both curves is comparable in the first half
of the evacuation. Later, the ellipses become smaller for the runs
with cooperative behaviour. In that case, the stronger aggregation of
the particularly cooperative group members is represented by a closer configuration of the participants.

As a consequence, larger groups seem often to require more space. This
effect is mitigated by including cooperative behaviour. Cooperative
groups cover less space compared to groups of the same size but with
normal behaviour. Accordingly, the stronger aggregation of cooperative
groups that was observed in the spatio-temporal diagrams reflects that
the participants stay closer than usual during the evacuation.

\subsubsection{Orientation of social groups}
\label{SSS3.2.3:GyrTensor}

\begin{figure}
\centering
\includegraphics[width=0.6\textwidth]{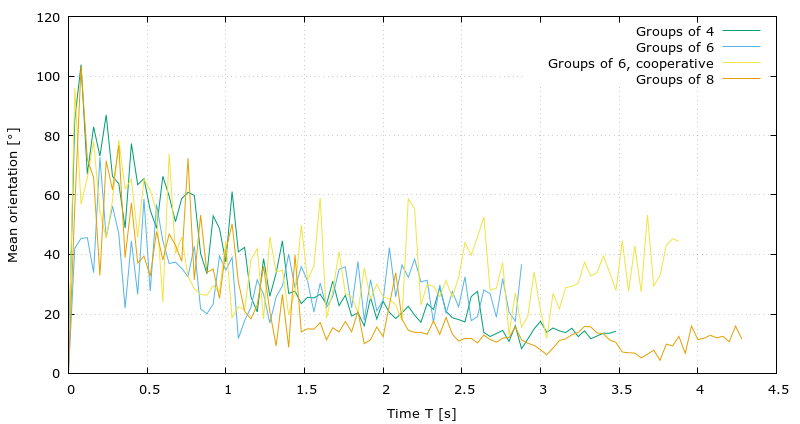}
\includegraphics[width=0.6\textwidth]{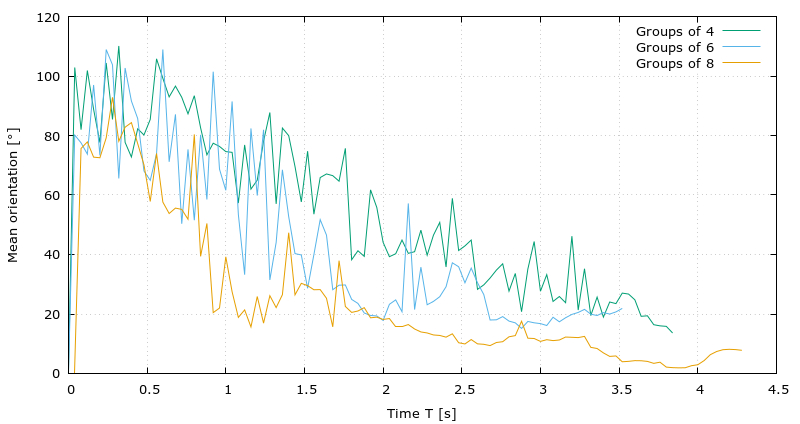}
%\subfigure[``GymBay"]{\includegraphics[width=0.49\textwidth]{Figures/Fig12-MeanThGymBay-oT.png}}
%\subfigure[``WDG"]{\includegraphics[width=0.49\textwidth]{Figures/Fig13-MeanThWDG-oT.png}}
\caption{The mean orientation of the centre of mass decreases for 
  all runs during the evacuation (Top: ``GymBay"; Bottom: ``WDG"). The
  groups move in a narrow region around the middle axis of the room.}
\label{Fig9:MeanOr}
\end{figure}

The orientation of the social groups can be determined in different
ways. First, the orientation of the centre of mass can be calculated.
It is defined as the angle between the velocity of the centre of mass
and an axis of reference \cite{Ballerini2008}. In our case this is the
room axis $\vec{e}_y$ that points into the room perpendicular to the exit door. Therefore, the orientation angle of the social group can be determined using the scalar product:

\begin{equation}
\Theta(t)=\pi-\arccos\left(\frac{v_y(t)}{v(t)}\right)
\end{equation}

where $v_y(t)$ is the component of the centre of mass' velocity in the direction of $\vec{e}_y$ and $v(t))$ is the absolute vlaue of the velocity at time $t$. This angle represents the orientation of the movement of the entire
group. The averaged values are shown in Fig.~\ref{Fig9:MeanOr} for
``GymBay" and ``WDG". In both cases, the angle decreases (after a
short starting phase) and flattens at low values between $0^\circ$ and
$40^\circ$. During the evacuation the social groups move in a narrow
cone around the door. The orientation is comparable for all runs in
``GymBay", whereas for the runs of ``WDG" the angle decreases with
increasing group size. In these cases, the cone in which the
pedestrians move is even narrower for larger groups.

%%%%%%%%%%%%%%%%%%%%%%%%%%%%%%%%
% Suggested size: 1.5 or 2 column
%%%%%%%%%%%%%%%%%%%%%%%%%%%%%%%%

\begin{figure}[h]
\centering
\includegraphics[width=0.6\textwidth]{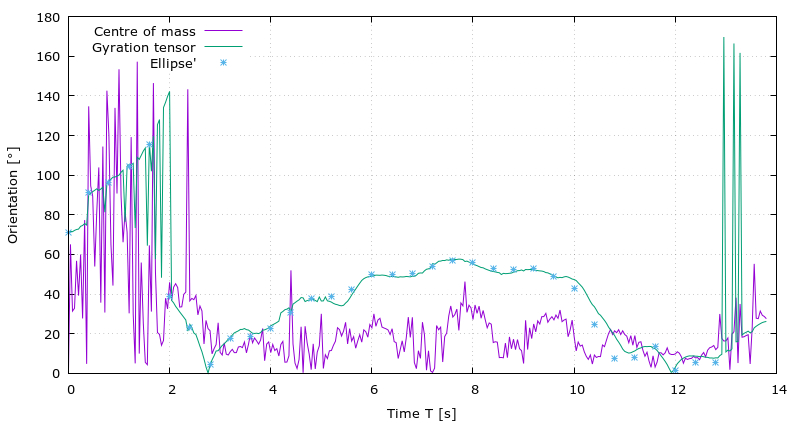}
\includegraphics[width=0.6\textwidth]{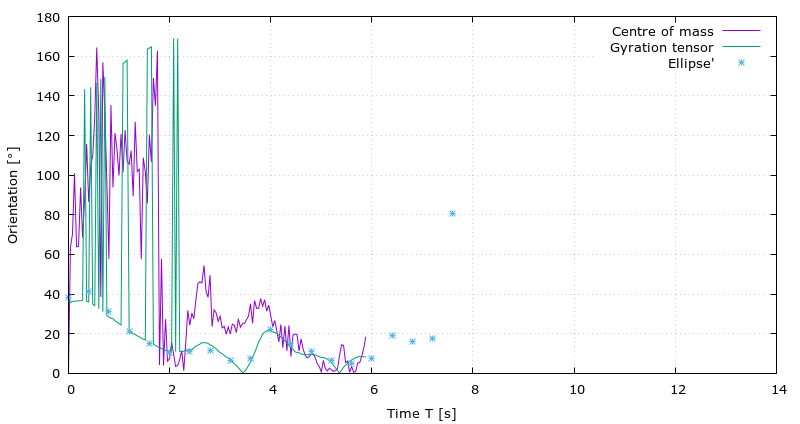}
%\subfigure[``GymBay"]{\includegraphics[width=0.49\textwidth]{Figures/Fig14-CompOrientGymBay-oT.png}}
%\subfigure[``WDG"]{\includegraphics[width=0.49\textwidth]{Figures/Fig15-CompOrientWDG-oT.png}}
\caption{Comparison of the orientation of an exemplary group for 
  ``GymBay" (top) and ``WDG" (bottom). The orientation calculated via
  the gyration tensor and the ellipse's orientation coincide well. The
  group members often orientate along their direction of movement.}
\label{Fig10:CompOr}
\end{figure}

Another way to describe the orientation of the social groups is to
determine the orientation of the ellipses \cite{Moshtag2006}. It is defined as the angle between the major axis of the ellipses and the room axis $\vec{e_y}$. Here,
the differences between the orientation curves of single groups are
rather large, especially for groups of four. Hence, it is not sensible
to calculate an averaged value. These differences arise from the
special configuration four pedestrians can acquire. Four pupils can
order in a square-like shape in contrast to groups with six or eight
persons. While moving in this square configuration even small shifts
of single persons can lead to considerable shifts of the ellipse's
orientation. Nevertheless, the orientation of the ellipse can be
compared to the orientation of the centre of mass for each single
group. A good accordance of both orientations would mean that the
group members order along the direction of movement. Indeed, many of
the orientations match at least temporarily as shown in
Fig.~\ref{Fig10:CompOr}. Therefore, pedestrians in social groups seem
to order along their general direction of motion.

The groups' orientation can also be calculated by a third approach
based on the gyration tensor \cite{Maurer2016}
\begin{equation}
G_{ij}(t)=\frac{1}{N_m}\sum_k^{N_m}(x_{k,i}(t)-x_{\text{CoM},i}(t))\cdot(x_{k,j}(t)-x_{\text{CoM},j}(t))
\end{equation}
where $N_m$ is again the number of group members, $\vec{x}_k$ the position of the group member $k$ and $\vec{x}_\text{CoM}$ the position of the centre of mass at time $t$. The orientation of the group is then defined as the eigenvector to the largest eigenvalue in our case. This approach has e.g. been used to determine the
orientation of red blood cells \cite{Maurer2016}. The gyration tensor is calculated using the positions of the
pedestrians in the group. Therefore, this orientation of the single social group is directly determined by the positioning of the
group members. It matches the orientations of the ellipses well for
the single groups at least temporarily. This indicates that the approximation of social
groups as ellipses is an appropriate choice to describe their internal
dynamics.

%%%%%%%%%%%%%%%%%%%%%%%%%%%%%%%%
% Suggested size: 1.5 or 2 column
%%%%%%%%%%%%%%%%%%%%%%%%%%%%%%%%

Overall, the social groups seem to orientate along their direction of
motion that proceeds within a narrow cone around the door.

%%%%%%%%%%%%%%%%%%%%%%%%%%%%%%%%%%%%%%%%%%%%%%%%%%%%%%%%%%%%%%%%%%%%%%%%

\section{Conclusion}
\label{S4:Conclusion}

The presence of social groups in pedestrian crowds and their
characteristics have been observed experimentally for several years
now. In order to analyse the influence of social groups on evacuation
scenarios we have conducted an empirical study with pupils. The
influence of social groups is investigated using data of
youths that evacuate in social groups of different sizes and different
intra-group interactions.

The evaluation of evacuation times shows that the presence of social
groups can have an advantageous influence on the evacuation process.
If the pedestrians form a queue in front of the door, the evacuation
is faster compared to a broad distribution of the pedestrians around the exit. The
reduction of conflicts arising from this ordering within the groups
may accelerate the whole evacuation process. In addition, evacuations
were faster for pedestrians that move compactly, especially at an early
stage of evacuation.

We have found several characteristics of the dynamics of the
  social groups during the evacuation process. Social groups
accelerate at the beginning of the evacuation but reach a constant
velocity soon. There are no distinct differences in group velocity
that might explain differences in evacuation times between the runs
with pairs and large groups. During the evacuation, social groups can
well be approximated by ellipses which typically have a prolately
elongated shape. In addition, their orientation shows that groups move
in a narrow cone around the exit.  Following the orientation of the
ellipses, members of social groups order along their direction of
motion. Groups of large sizes have an higher space requirement per person, especially for groups of eight. Therefore, faster
evacuations for larger group sizes do not arise from denser
configurations of the group members. The comparison of the
orientations also shows that the approximation of groups as ellipses
supplies reasonable results.

Explicit cooperative behaviour can inhibit an advantageous impact of
groups. The pedestrians concentrate on their own group members and pay
less attention to other pedestrians. Therefore, the participants evacuate in
an intermittent way separated into several bursts. The stronger
aggregation within the social group is also reflected in a smaller
space requirement for cooperative behaviour compared to same group
size with normal behaviour. In addition, groups with cooperative
behaviour reach a lower constant velocity level. All this factors lead
to a slower evacuation.

Summarizing, our experiments have shown that the presence of social
groups can have a substantial impact on evacuation scenarios. We have
introduced ``group parameters" which allow to quantify the dynamics of
groups and their members.

%%%%%%%%%%%%%%%%%%%%%%%%%%%%%%%%%%%%%%%%%%%%%%%%%%%%%%%%%%%%%%%%%%%%%%%%

\section*{Acknowledgements}
\label{S5:Acknowledgements}

We thank the teams of the Universities of Wuppertal and Cologne and
the Forschungszentrum J\"ulich for their support with the experiments,
especially Maik Boltes for providing the trajectory data. We also
thank the students and teachers of Gymnasium Bayreuther Stra{\ss}e and
Wilhelm-D\"orpfeld-Gymnasium for their participation. Financial
support by Deutsche Forschungsgemeinschaft (DFG) under grant SCHA
636/9-1 and Bonn-Cologne Graduate School of Physics and Astronomy
(BCGS) is gratefully acknowledged.

\appendix

\section{Details of experimental runs}
\label{AS1:Details}

\subsection{List of experiments}
\label{ASS1.1:List}

\begin{table}[h]
\centering
\fbox{
\begin{tabular}{c|c|c|c|c}
Run							&	\multicolumn{2}{c}{Number of participants}	&	\multicolumn{2}{c}{Male participants}\\
(Youths ($\sim$~16 years))							&	``GymBay"	&	``WDG"	&	``GymBay"	&	``WDG"\\
(No leader-follower relationship)\\ \hline
\\
Individuals 				& 	-			&	41		&		-		&	29\\
Pairs						&	46			&	42		&		21		&	30\\
Groups of Four				&	44			&	40		&		20		&	29\\
Groups of Six				&	42			&	42		&		17		&	30\\
Groups of Six, cooperative	&	42			&	 - 		&		17		&	- \\
Groups of Eight				&	40			&	40		&		15		&	29\\
\end{tabular}
}
\caption{Experimental runs that are used for the analysis of the influence of social groups with the respective number of participants.}
\label{T1:UsedData}
\end{table}

The proportion of male participants in the crowd was estimated from the video recordings.

\subsection{Distribution of body height in the experiments}
\label{ASS1.2:BodyHeight}

In order to classify the students' body heights it was measured before the experiments. Afterwards, the interval between the maximum and minimum body height of the entire group was divided by the number of available cap colours. In the Gase of ``GymBay", the youths were separated into five different body height intervals, for ``WDG", there were four categories for the young adults. For each body height interval the middle value was approximated and is stated in the tables below. It represents approximately the averaged body height of the persons belonging to the respective body height interval.

\begin{table}[h]
\centering
\fbox{
\begin{tabular}{c|c|c|c|c|c}
Run							&	161.6 cm	&	168.2	&	176.1 cm	&	182.7 cm	&	189.1 cm\\
\\ \hline
\\
Pairs						&	7			&	6		&	15			&	8			&	10\\
Groups of Four				&	7			&	6		&	15			&	8			&	8\\
Groups of Six				&	7			&	6		&	15			&	8			&	6\\
Groups of Six, cooperative	&	7			&	6 		&	15			&	8			&	6\\
Groups of Eight				&	7			&	6		&	15			&	8			&	4\\

\end{tabular}
}
\caption{Number of participants in the five body height intervals, respectively, for the experiments at ``GymBay"}
\label{T2:BodyHeightGymBay}
\end{table}

\begin{table}[h]
\centering
\fbox{
\begin{tabular}{c|c|c|c|c}
Run				&	156.6 cm	&	166.4 cm	&	179.3 cm	&	191.7 cm\\
\\ \hline
\\
Individuals		&	2			&	10			&	23			&	6\\
Pairs 			&	2			&	11			&	23			&	6\\
Groups of Four	&	1			&	11			&	22			&	6\\
Groups of Six	&	2			&	11 			&	23			&	6\\
Groups of Eight	&	2			&	11			&	21			&	6\\

\end{tabular}
}
\caption{Number of participants in the four body height intervals for the youths, respectively, for the experiments at ``WDG"}
\label{T3:BodyHeightWDG}
\end{table}

\subsection{Instructions during the experiments}
\label{ASS1.3:Instructions}

Instructions for the experiments at ``GymBay":
\begin{itemize}
	\item \textit{Pairs:}	Try to stay together with your group partner and to leave the room as fast as possible. There is no leader.
	\item \textit{Groups of four:}	Tray to stay together. Imagine that you have visited a concert with friends and you don't want to loose them while leaving the room.
	\item \textit{Groups of six:}	The same as before: try to stay together in your groups and not to be split up.
	\item \textit{Groups of six with cooperative behaviour:}	The groups should be close even at the beginning of the evacuation. Try more to stay together and let another group go first if there is the risk that your own group would be split up otherwise. Imagine that you're visiting a place you don't know and that the group members have to find the way all together.
	\item \textit{Groups of eight:}	The same as before, try to stay together.
\end{itemize}

Instructions for the experiments at ``WDG":
\begin{itemize}
	\item \textit{Individual:}	Try to leave the room as fast as possible without running or scrambling. Imagine that you're leaving your classroom.
	\item \textit{Pairs:} Try to stay together and to leave the room quickly. There are equal partners and no leaders.
	\item \textit{Groups of four:}	Try to stay together in your groups of four without scrambling each other.
	\item \textit{Groups of six:}	The same as before, try to stay together.
	\item \textit{Groups of eight:}	Try to stay together without scrambling each other.	
\end{itemize}



\section*{References}
\label{References}

\end{document}